\documentclass[traditabstract]{aa} 
\usepackage{longtable}
\usepackage{latexsym}
\usepackage{amssymb}
\usepackage{lscape}
\usepackage[]{natbib}

\usepackage{color}

\usepackage{graphicx}
\usepackage{txfonts}
\def\lsim{\mathrel{\rlap{\lower 3pt \hbox{$\sim$}} \raise 2.0pt \hbox{$<$}}}
\def\gsim{\mathrel{\rlap{\lower 3pt \hbox{$\sim$}} \raise 2.0pt \hbox{$>$}}}
\title{A 85 kpc H$\alpha$ tail behind 2MASX J11443212+2006238 in A1367\thanks{based on observations taken at the San Pedro Martir telescope belonging to the Mexican National Observatory (OAN)}}
\subtitle{}
\author{G. Gavazzi \inst{1} 
\and G. Consolandi \inst{1}                                                                
\and M. Yagi \inst{2,3}                            
\and M. Yoshida \inst{4}
}
\authorrunning{G. Gavazzi  et al.}
\titlerunning{A 85 kpc H$\alpha$ tail in A1367}
\institute{Universit\`a degli Studi di Milano-Bicocca, Piazza della Scienza 3, 20126 Milano, Italy\\
\email {giuseppe.gavazzi@mib.infn.it}
\and
Optical and Infrared Astronomy Division, National Astronomical Observatory of Japan, Mitaka, Tokyo, 181-8588, Japan\\
\email {yagi.masafumi@nao.ac.jp}
\and
Graduate School of Science and Engineering, Hosei University, 3-7-2, Kajinocho, Koganei, Tokyo, 184-8584 Japan
\and
Subaru Telescope, National Astronomical Observatory of Japan,
National Institutes of Natural Sciences,
650 A'ohoku Place, Hilo, Hawaii 96720, USA
}
\begin{document}

\date{Received; accepted}
\abstract{We report the detection of an H$\alpha$ trail of 85 kpc projected length behind galaxy 2MASX J11443212+2006238 in the nearby cluster of galaxies
Abell 1367. This galaxy was discovered to possess an extended component in earlier, deeper H$\alpha$ observations carried out with the Subaru telescope.
However, lying at the border of the Subaru field, the extended H$\alpha$ tail was cut out, preventing the determination of its full  extent. We fully map this extent here, albeit the shallower exposure. 
 } 
\maketitle
%

\section{Introduction}

A 1.5 $\rm deg^2$ region of the nearby cluster of galaxies Abell 1367 ($z\sim0.0217$) was recently surveyed with deep
H$\alpha$ observations using the Subaru telescope (Yagi et al. 2017). 
These observations, at the limiting surface brightness of $2.5\times 10^{-18}~\rm erg ~cm^{-2}~sec^{-1}~arcsec^{-2}$,
revealed the presence of H$\alpha$ tails behind ten out of the cluster's 26 late-type galaxies (LTG) that were surveyed.
This indicates that, when observed with sufficiently deep observations, 
approximately 40 \% of all LTGs in this cluster reveal an associated
extended trail of H$\alpha$, in agreement with the frequency obtained in the Coma cluster by Yagi et al. (2010).
This evidence strengthens previous suggestions that a massive infall of gas-rich, star forming galaxies 
(100-400 galaxies per Gyr, Adami et al. 2005, Boselli et al 2008; Gavazzi et al 2013, 2013b) 
is occurring at the present epoch onto rich clusters of galaxies such as Coma  and Virgo. 
This estimate derives from the combined evidence that ionized tails arise from the ram pressure stripping (Gunn \& Gott, 1972) 
of galaxies crossing the intracluster medium (ICM) at high speed for the first time, and that the gas ablation produced by such interaction
proceeds on timescales as short as 100 Myr (Boselli \& Gavazzi, 2006, 2014).\\
\begin{figure*}
\centering
\includegraphics[width=19.cm]{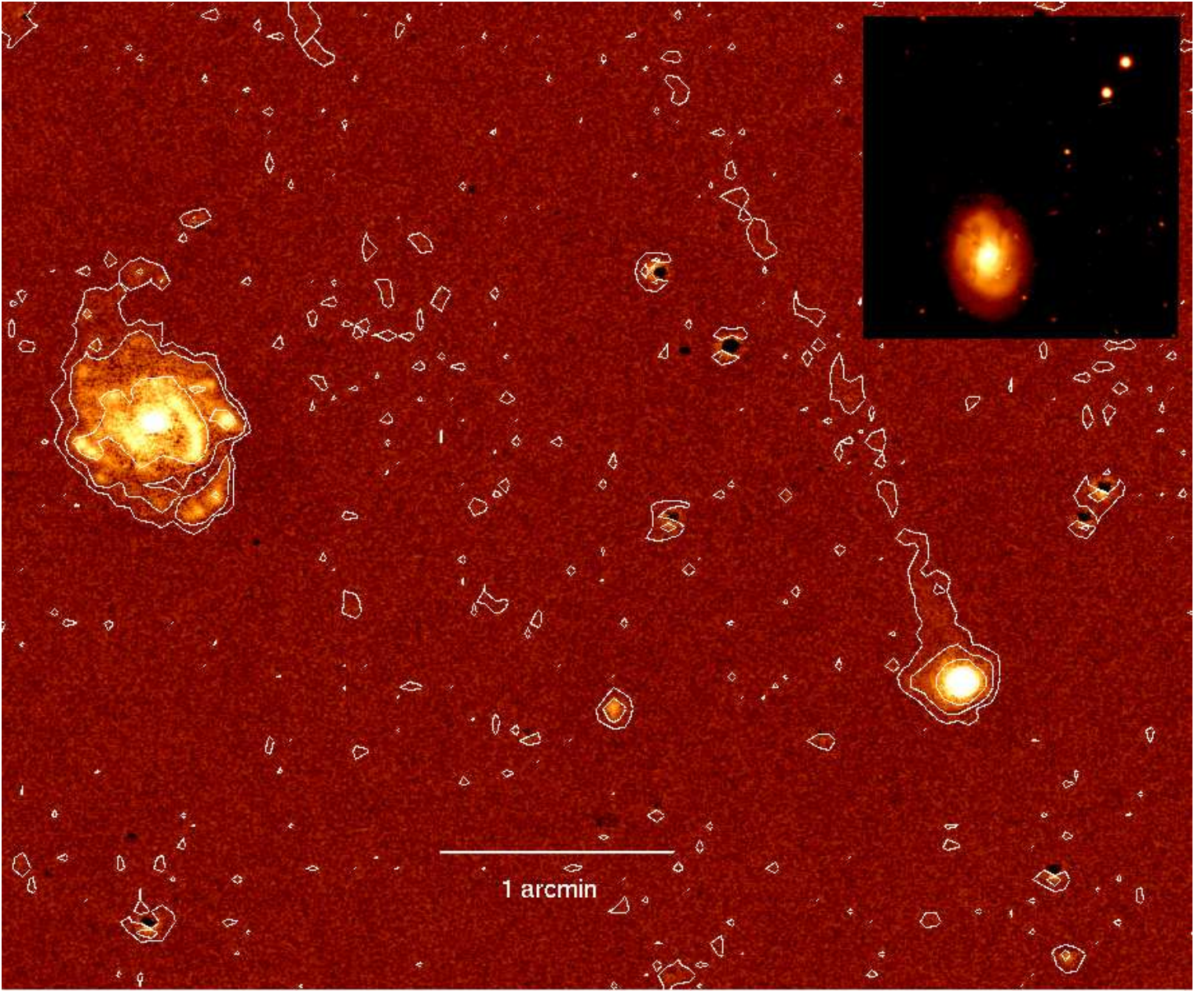}
\caption{Gray-scale representation of the H$\alpha$ emission from CGCG 97-121 (left) and 2MASX J11443212+2006238 (right),
with levels chosen to highlight the bright part of the emission. Smoothed (by 2.8 arcsec) contours of the H$\alpha$-NET image show the low-brightness,
extended tail behind 2MASX J11443212+2006238 whose maximum projected length is 85 kpc. North is at the top and East to the left.}\
\label{NET}  
\end{figure*}
\begin{table*}
\caption{Observational parameters of the two target galaxies.}                                                                   
\centering                                                                                        
\begin{tabular}{l c c c c c c}                                                                    
\hline\hline                                                                                      
  Object  &    RA                &    Dec       &    cz           & $M_*$      &    H$\alpha$ Flux             & H$\alpha$ E.W. \\                  
          &  (hh m s)            & ($^o$ ' ")   & $\rm km~s^{-1}$ & $M_{\odot}$& $\rm erg ~cm^{-2}~sec^{-1}$   &  \AA           \\                 
\hline                                                                                            
 2MASX J11443212+2006238   &    11 44 32.1      &   20 06 24  &  7214  & 9.87  & $ 6   \times 10^{-14}$  (4.8   $\times 10^{-15}$) & 32.3    \\
 CGCG 97-121               &    11 44 47.0      &   20 07 30  &  6571  & 10.57 & $ 8.5 \times 10^{-14}$ & 6.8      \\     
  
\hline 
\end{tabular}                                                                                     
\label{tab1}                                                                                      
\end{table*}

The galaxy 2MASX  J11443212+2006238 lies very close to the edge of the 
Subaru field (see Fig. 9 of Yagi et al. 2017), preventing a robust measurement of the extended  H$\alpha$ flux and of its total length.
Guided by these observations, we decided to devote two entire (moon free, clear but not photometric) nights of observations at the 2.1m telescope of the San Pedro Martir observatory 
to the field containing 2MASX J11443212+2006238 and another cluster member: CGCG 97-121 (Zwicky et al.  1961-68). In spite of the three-times-shallower data obtained
with our smaller telescope, we could detect the full extent of the H$\alpha$ emission trailing behind 2MASX J11443212+2006238. 
\section{Observations}
\label{Obs}
On the nights of April 24 and 25, 2017, we used the 2.1m telescope at San Pedro Martir  to repeatedly observe a 5 x 5 $\rm arcmin^2$ region of the
nearby cluster of galaxies Abell 1367.
 We used a narrow (80 \AA) band filter centered at 6723 \AA ~to detect the H$\alpha$ emission (at the mean redshift $z\sim0.0217$ of the cluster) 
and a broad band $r$ (Gunn) filter (effective $\lambda$ 6231 \AA, $\Delta\lambda \sim 1200 ~\AA$) to recover the continuum emission.

The field, which contains both 2MASX J11443212+2006238 and 
CGCG 97 121, was observed with 33 independent pointings of 900 seconds with the ON-band filter,
and with 25 pointings of 180 seconds with the broad band (OFF-band) $r$ filter. After flat-fielding, the aligned observations were combined 
into a final ON-band frame totalling 8.25 hours exposure, and an OFF-band frame of 1.25 hours exposure (details on the reduction procedures of H$\alpha$
observations can be found in Gavazzi et al. 2012).
To calibrate the data, which were observed in clear sky conditions, we used the Sloan Digital Sky Survey (SDSS)
nuclear fiber spectrum of 2MASX J11443212+2006238 from which we
measure an H$\alpha$ point-like flux (within the central 3 arcsec) of $2.45 \times 10^{-14}$ $\rm erg ~cm^{-2}~sec^{-1} \AA^{-1}$.
Assuming this flux within the 3" central aperture of the galaxy, we derive an effective zero point for our observations of -15.29 $\rm erg ~cm^{-2}~sec^{-1}$ 
this should be compared with -15.54 obtained  in the previous photometric nights
and we estimate a limiting surface brightness at H$\alpha  ~\approx 7.9\times 10^{-18}~\rm erg ~cm^{-2}~sec^{-1}~arcsec^{-2}$.

\section{Results}
In Table \ref{tab1} we list the celestial coordinates, redshift, total stellar mass, and total equivalent width (E.W.) and flux separately for 2MASX J11443212+2006238, 
for the diffuse trailing material, and for CGCG 97 121. 
The flux of the diffuse gas of 2MASX J11443212+20062 (in parenthesis) is measured in a polygonal aperture that fits the last contour level shown in Fig. \ref{NET}. 
Stellar masses were derived from the $i$-band luminosity and the $g-i$ color according to Zibetti et al (2009), assuming a Chabrier initial mass function (IMF) (Chabrier, 2003).
Fig. \ref{NET} shows the continuum subtracted H$\alpha$ image of our field smoothed by 2.8 arcsec, from which we measured the extent of the diffuse flux associated to 2MASX J11443212+2006238. 
Assuming a distance from Abell 1367 of 95 Mpc, the tail extends in the NE direction for $\approx 85\rm ~kpc$. Such a distance can be covered in $\approx 85$ Myr
by a galaxy travelling at approximately 1000 $\rm km~s^{-1}$, which is the typical velocity of galaxies in clusters.
Given that this length is typical in Abell 1367 (the tail length ranges from 15 to 230 kpc in this cluster, Yagi et al. 2017), 85-100 Myr
can be assumed as the typical time during which the tails remain visible. This time is longer than the typical electron recombination time, recently estimated as 0.2 Myr for UGC 6697 
(Consolandi et al. 2017) and even shorter for ESO137-001 (Fossati et al. 2016), 
suggesting the existence of some mechanism capable of keeping the gas ionized for a longer period of time in the wakes of the extended tails.
However, 100 Myr is a rather short period of time in cosmic history. 
This suggests that the ram pressure stripping phenomenon occurs during the first pericenter passage over a short period of time, consistent 
with the evidence of truncated star formation at the southern side
(opposite to the tail) of 2MASX J11443212+20062 (Yagi et al 2017), which could indicate quenching timescales as short as 100 Myr (see Boselli et al. 2006, 2016a, 2016b).  
The very existence of 11 tails of similar typical length allows us to roughly estimate the infall rate of galaxies in A1367 as 130 per Gyr,
consistent with other existing estimates for Coma and Virgo.
%
\begin{acknowledgements}
This research has made use of the GOLDmine database (Gavazzi et
al. 2003, 2014b) and of the NASA/IPAC Extragalactic Database (NED), which is
operated by the Jet Propulsion Laboratory, California Institute of Technology, under contract with the National Aeronautics and Space Administration.  
\end{acknowledgements}

\end{document}